%% file: ms.tex
\documentclass[preprint,tightenlines,a4paper,aps,prd,superscriptaddress,preprintnumbers,amssymb,amsmath,alignedleftspaceno]{revtex4-2}
\pdfoutput=1
\usepackage[utf8]{inputenc}
\usepackage[T1]{fontenc}
\usepackage[USenglish]{babel}
\usepackage{csquotes}
\usepackage[all]{foreign}
\defasforeign[ansaetze]{ansätze}
\renewcommand{\cf}{\xperiodafter{{\foreignabbrfont{cf}}}}

\usepackage{tikz}
\usepackage{braket}

\newcommand{\pd}[2]
{
    \frac{\partial #1}{\partial #2}
}
\usepackage[pdfusetitle,pdfauthor={Miguel Salg},bookmarksopen=true,bookmarksopenlevel=1]{hyperref}
\usepackage[english]{cleveref}
\usepackage{graphicx}
\usepackage{tikz}
\usepackage{longtable}
\usepackage[protrusion=true,expansion,tracking,kerning]{microtype}
\microtypecontext{spacing=nonfrench}
\newdimen\defaultaddspace
\usepackage{xstring}
\newcommand{\num}[1]
{%
    \IfDecimal{#1}%
    {%
        #1%
    }%
    {%
        \IfSubStr{#1}{e}%
        {%
            \StrCut{#1}{e}{\mantissa}{\exponent}%
            \IfDecimal{\mantissa}{}{\errmessage{Argument of \noexpand\num is not a valid number}}%
            \IfInteger{\exponent}{}{\errmessage{Argument of \noexpand\num is not a valid number}}%
            \ensuremath{\mantissa \times 10^{\number\exponent\relax}}%
        }%
        {%
            \IfSubStr{#1}{E}%
            {%
                \StrCut{#1}{E}{\mantissa}{\exponent}%
                \IfDecimal{\mantissa}{}{\errmessage{Argument of \noexpand\num is not a valid number}}%
                \IfInteger{\exponent}{}{\errmessage{Argument of \noexpand\num is not a valid number}}%
                \ensuremath{\mantissa \times 10^{\number\exponent\relax}}%
            }%
            {%
                \errmessage{Argument of \noexpand\num is not a valid number}%
            }%
        }%
    }%
}%

\begin{document}
\title{Zemach and Friar radii of the proton and neutron from lattice QCD}
\author{Dalibor Djukanovic}
 \affiliation{Helmholtz Institute Mainz, Staudingerweg 18, 55128 Mainz, Germany}
 \affiliation{GSI Helmholtzzentrum für Schwerionenforschung, 64291 Darmstadt, Germany}
\author{Georg von Hippel}
 \affiliation{\texorpdfstring{PRISMA${}^+$}{PRISMA+} Cluster of Excellence and Institute for Nuclear Physics, Johannes Gutenberg University Mainz, Johann-Joachim-Becher-Weg 45, 55128 Mainz, Germany}
\author{Harvey B. Meyer}
 \affiliation{Helmholtz Institute Mainz, Staudingerweg 18, 55128 Mainz, Germany}
 \affiliation{\texorpdfstring{PRISMA${}^+$}{PRISMA+} Cluster of Excellence and Institute for Nuclear Physics, Johannes Gutenberg University Mainz, Johann-Joachim-Becher-Weg 45, 55128 Mainz, Germany}
\author{Konstantin Ottnad}
 \affiliation{\texorpdfstring{PRISMA${}^+$}{PRISMA+} Cluster of Excellence and Institute for Nuclear Physics, Johannes Gutenberg University Mainz, Johann-Joachim-Becher-Weg 45, 55128 Mainz, Germany}
\author{Miguel Salg}
 \email{msalg@uni-mainz.de}
 \affiliation{\texorpdfstring{PRISMA${}^+$}{PRISMA+} Cluster of Excellence and Institute for Nuclear Physics, Johannes Gutenberg University Mainz, Johann-Joachim-Becher-Weg 45, 55128 Mainz, Germany}
\author{Hartmut Wittig}
 \affiliation{Helmholtz Institute Mainz, Staudingerweg 18, 55128 Mainz, Germany}
 \affiliation{\texorpdfstring{PRISMA${}^+$}{PRISMA+} Cluster of Excellence and Institute for Nuclear Physics, Johannes Gutenberg University Mainz, Johann-Joachim-Becher-Weg 45, 55128 Mainz, Germany}
\begin{abstract}
    \input{content/abstract}
\end{abstract}
\date{\today}
\preprint{MITP-23-055}

\maketitle
\newpage
\input{content/introduction}
\input{content/setup}
\input{content/bchpt_fits}
\input{content/extrapolation_integration}
\input{content/model_average}
\input{content/conclusions}
\input{content/acknowledgments}

\bibliography{literature.bib}

\appendix*
\input{content/appendix}
\end{document}

%% file: content/abstract.tex
We present the first lattice-QCD result for the Zemach and Friar radii of the proton and neutron.
Our calculation includes both quark-connected and -disconnected diagrams and assesses all sources of systematic uncertainties arising from excited-state contributions, finite-volume effects and the continuum extrapolation.
At the physical point, we obtain for the proton $r_Z^p = ( 1.013 \pm 0.010\ (\mathrm{stat}) \pm 0.012\ (\mathrm{syst}) )~\mathrm{fm}$ and $r_F^p = ( 1.301 \pm 0.012\ (\mathrm{stat}) \pm 0.014\ (\mathrm{syst}) )~\mathrm{fm}$.
These numbers suggest small values of the Zemach and Friar radii of the proton, but are compatible with most of the experimental studies.

%% file: content/introduction.tex
\section{Introduction}
The most accurate determination of the proton's electric (charge) radius is derived from the measurement of the Lamb shift in muonic hydrogen spectroscopy \cite{Pohl2010,Antognini2013}.
This result exhibits a large tension with some $ep$-scattering experiments \cite{Bernauer2014,Mihovilovic2021}, which is known as the \enquote{proton radius puzzle}.

To infer the electric radius from the observed Lamb shift, higher-order nuclear structure-contributions need to be subtracted.
The leading contribution is the two-photon exchange \cite{Pachucki2024}, the dominant, elastic part of which depends on the third Zemach moment of the proton \cite{Friar1979,Pachucki1996,Pachucki1999},
\begin{align}
    \langle r_E^3 \rangle_{(2)}^p &= \frac{48}{\pi} \int_0^\infty \frac{dQ}{Q^4} \left[(G_E^p(Q^2))^2 - (G_E^p(0))^2 - \left.\pd{(G_E^p(Q^2))^2}{Q^2}\right|_{Q^2 = 0} Q^2\right] \nonumber \\
    &= \frac{24}{\pi} \int_0^\infty \frac{dQ^2}{(Q^2)^{5/2}} \left[(G_E^p(Q^2))^2 - 1 + \frac{1}{3}\langle r_E^2 \rangle^p Q^2\right] .
    \label{eq:third_Zemach_moment}
\end{align}
The associated radius is known as the Friar radius of the proton,
\begin{equation}
    r_F^p = \sqrt[3]{\langle r_E^3 \rangle_{(2)}^p} .
    \label{eq:Friar_radius}
\end{equation}
A very large Friar radius was once suggested~\cite{DeRujula2010} as a possible solution to the proton radius puzzle.
For this purpose, however, the Friar radius would need to be so large that the expansion in radii would break down \cite{Distler2011,Hagelstein2015}.

While the traditional proton radius puzzle awaits its final resolution, the goal of reaching a consistent picture of all the fundamental electromagnetic properties of the nucleon has attained a new prominence.
Historically, data-driven dispersive approaches had found values of the electric radii of the proton consistent with the lower value of muonic-atom spectroscopy measurements \cite{Mergell1996,Belushkin2007}.
For the magnetic properties, a tension between dispersive approaches \cite{Lin2022} and $z$-expansion results \cite{Borah2020} appeared, \ie a separate puzzle beclouds the magnetic properties of the proton.
Underlining the importance of the magnetic properties of the proton, several experiments are under way to measure these from spectroscopy on (muonic) hydrogen \cite{Sato2014,Pizzolotto2020,Pizzolotto2021,Amaro2022}.
This can be achieved by measuring, in addition to the Lamb shift, the hyperfine splitting (HFS) in either electronic or muonic hydrogen, which is caused by the magnetic spin-spin interaction between the nucleus and the orbiting lepton.
The influence of the electromagnetic structure of the nucleus on the HFS is particularly pronounced for the $S$-states, since the $S$-state wavefunction has a large overlap with the nucleus.

The leading-order proton-structure contribution to the $S$-state HFS of hydrogen depends on the Zemach radius of the proton \cite{Zemach1956,Pachucki1996},
\begin{align}
    r_Z^p &= -\frac{4}{\pi} \int_0^\infty \frac{dQ}{Q^2} \left[\frac{G_E^p(Q^2) G_M^p(Q^2)}{\mu_M^p} - \frac{G_E^p(0) G_M^p(0)}{\mu_M^p}\right] \nonumber \\
    &= -\frac{2}{\pi} \int_0^\infty \frac{dQ^2}{(Q^2)^{3/2}} \left[\frac{G_E^p(Q^2) G_M^p(Q^2)}{\mu_M^p} - 1\right] .
    \label{eq:Zemach_radius}
\end{align}
Having a first-principles prediction of the Zemach radius  prior to the experimental measurement of the ground-state ($1S$) HFS in muonic hydrogen with ppm precision \cite{Sato2014,Pizzolotto2020,Pizzolotto2021,Amaro2022}, from which the Zemach radius could be extracted with sub-percent uncertainty, is highly desirable.
Beyond helping in narrowing down the frequency search range, such a prediction allows for a crucial consistency check.
We note that the interpretation of the experimental HFS results relies on theoretical input
for the proton-polarizability effect, where a discrepancy has emerged between data-driven approaches and baryon chiral perturbation theory \cite{Hagelstein2023}.
Eventually, combining precise HFS measurements in electronic and muonic hydrogen,
the proton polarizability can be determined from those as well and compared to theory \cite{Antognini2022}.

In this letter we present the first lattice-QCD calculation of the Zemach and Friar radii, building on our results for the electromagnetic radii of the proton and neutron \cite{Djukanovic2023,Djukanovic2023a}.
Our results for the Zemach and Friar radii of the proton have a total precision of 1.5~\%, and are well compatible with most of the experimental determinations \cite{Volotka2005,Distler2011,Antognini2013,Borah2020,Lin2022,Hagelstein2023}.

%% file: content/setup.tex
\section{Lattice setup}
In order to compute the Zemach and Friar radii of the proton and neutron, we need, according to \cref{eq:third_Zemach_moment,eq:Zemach_radius}, information on their electric and magnetic form factors.
For our lattice determination of the latter, we employ a set of lattice ensembles with $N_f = 2 + 1$ dynamical flavors of non-perturbatively $\mathcal{O}(a)$-improved Wilson fermions \cite{Sheikholeslami1985,Bulava2013}, using the tree-level improved Lüscher-Weisz gluon action \cite{Luescher1985} and twisted-mass reweighting \cite{Luescher2013,Palombi2009}, which have been generated as part of the Coordinated Lattice Simulations (CLS) effort \cite{Bruno2015}.
The ensembles entering our analysis are listed in \cref{tab:ensembles} and cover four lattice spacings $a \in [0.050, 0.086]~\mathrm{fm}$ as well as several pion masses down to slightly below the physical one (E250).
The calculation of the reweighting factors and the correction of the strange-quark determinant are described in Refs.\@ \cite{Kuberski2024} and \cite{Mohler2020}, respectively.
We include the contributions from quark-connected as well as -disconnected diagrams.
For further details concerning the setup of the simulations, the calculation of our raw lattice observables, the extraction of the form factors, and the treatment of excited states, we refer to Ref.\@ \cite{Djukanovic2023}.

\begin{table}[htb]
    \caption{Overview of the ensembles used in this study. For further details, see table~I of Ref.\@ \cite{Djukanovic2023}.}
    \label{tab:ensembles}
    \begin{ruledtabular}
        \begin{tabular}{lccccc}
            ID                   & $\beta$ & $t_0^\mathrm{sym}/a^2$ & $T/a$ & $L/a$ & $M_\pi$ [MeV] \\ \hline
            C101                 & 3.40    & 2.860(11)              & 96    & 48    & 227           \\
            N101\footnotemark[1] & 3.40    & 2.860(11)              & 128   & 48    & 283           \\
            H105\footnotemark[1] & 3.40    & 2.860(11)              & 96    & 32    & 283           \\[\defaultaddspace]
            D450                 & 3.46    & 3.659(16)              & 128   & 64    & 218           \\
            N451\footnotemark[1] & 3.46    & 3.659(16)              & 128   & 48    & 289           \\[\defaultaddspace]
            E250                 & 3.55    & 5.164(18)              & 192   & 96    & 130           \\
            D200                 & 3.55    & 5.164(18)              & 128   & 64    & 207           \\
            N200\footnotemark[1] & 3.55    & 5.164(18)              & 128   & 48    & 281           \\
            S201\footnotemark[1] & 3.55    & 5.164(18)              & 128   & 32    & 295           \\[\defaultaddspace]
            E300                 & 3.70    & 8.595(29)              & 192   & 96    & 176           \\
            J303                 & 3.70    & 8.595(29)              & 192   & 64    & 266
        \end{tabular}
    \end{ruledtabular}
    \footnotetext[1]{These ensembles are not used in the final fits but only to constrain discretization and finite-volume effects.}
\end{table}

All dimensionful quantities are expressed in units of the gradient flow time $t_0$ \cite{Luescher2010}.
To this end, we use the numerical determination at the flavor-symmetric point, $t_0^\mathrm{sym}/a^2$, from Ref.\@ \cite{Bruno2017}.
Only our final results for the radii are converted to physical units using the FLAG estimate \cite{Aoki2021}
\begin{equation}
    \sqrt{t_{0, \mathrm{phys}}} = 0.14464(87)~\mathrm{fm}
    \label{eq:sqrt_t0_phys}
\end{equation}
for $N_f = 2 + 1$.

%% file: content/bchpt_fits.tex
\section{Fits to baryonic \texorpdfstring{$\chi$PT}{ChPT}}
\label{sec:bchpt_fits}
In Refs.\@ \cite{Djukanovic2023,Djukanovic2023a} we have combined the parametrization of the $Q^2$-dependence of the form factors with the extrapolation to the physical point ($M_\pi = M_{\pi, \mathrm{phys}}$, $a = 0$, $L = \infty$).
For this purpose, we have fitted our form factor data to the next-to-leading-order expressions resulting from covariant baryon chiral perturbation theory (B$\chi$PT) \cite{Bauer2012}.
While explicit $\Delta$ degrees of freedom are not considered in the fit, we include the contributions from the relevant vector mesons, as discussed in detail in Ref.\@ \cite{Djukanovic2023}.
For the physical pion mass we use the value in the isospin limit \cite{Aoki2014},
\begin{equation}
    M_{\pi, \mathrm{phys}} = M_{\pi, \mathrm{iso}} = 134.8(3)~\mathrm{MeV} ,
    \label{eq:m_pi_phys}
\end{equation}
so that in units of $t_0$, we employ $\sqrt{t_{0, \mathrm{phys}}} M_{\pi, \mathrm{phys}} = 0.09881(59)$.
Here, the uncertainty of $M_{\pi, \mathrm{iso}}$ in physical units is neglected since it is entirely subdominant compared to the uncertainty of the scale $\sqrt{t_{0, \mathrm{phys}}}$.

We perform several such fits, applying different cuts in the pion mass ($M_\pi \leq 0.23~\mathrm{GeV}$ and $M_\pi \leq 0.27~\mathrm{GeV}$) and the momentum transfer ($Q^2 \leq 0.3, \ldots, 0.6~\mathrm{GeV}^2$), and, at the same time, varying our model for the lattice-spacing and/or finite-volume dependence, in order to estimate the corresponding systematic uncertainties.
The aforementioned relatively strict cuts in $Q^2$ are required because the B$\chi$PT expansion, from which our fit formulae are derived, is only applicable for low momentum transfers.
By including the contributions from vector mesons, the range of validity of the resulting expressions can be extended \cite{Kubis2001,Schindler2005,Bauer2012}.
Nevertheless, as the heaviest vector meson we consider in the isovector channel is the $\rho$, momentum transfers larger than $M_\rho^2 \approx 0.6~\mathrm{GeV}^2$ cannot safely be described in this way.
For further technical details on our implementation of the B$\chi$PT fits, we refer to Ref.\@ \cite{Djukanovic2023}.

We have extensively crosschecked our excited-state analysis as well as the parametrization of the $Q^2$-dependence and the extrapolation to the physical point; for details, see Ref.\@ \cite{Djukanovic2023} and its appendices.

%% file: content/extrapolation_integration.tex
\section{Extrapolation of the form factors and integration}
Given that the Zemach radius and third Zemach moment are defined as integrals over all possible (spacelike) values of $Q^2$ [\cf \cref{eq:third_Zemach_moment,eq:Zemach_radius}], an extrapolation of the B$\chi$PT fits beyond their range of applicability is required if they are to be employed to parametrize the form factors.
For each model, we evaluate the B$\chi$PT formula for $G_E^{p,n}$ and $G_M^{p,n}$, using the low-energy constants as determined from the corresponding fit, at the physical point and at twenty evenly spaced points in $Q^2 \in (0, Q^2_\mathrm{cut}]$.
Here, $Q^2_\mathrm{cut}$ is the cut in the momentum transfer corresponding to the respective variation of the B$\chi$PT fit.

In the next step, we fit a model which obeys the large-$Q^2$ constraints on the form factors from perturbation theory \cite{Lepage1980} to these data points and their error estimates.
We note that the data points exhibit an extremely high correlation due to the way we generate them.
Taking these correlations into account when adjusting the extrapolation model would thus not be meaningful, and also technically challenging because the resulting covariance matrices are extremely badly conditioned.
To describe the $Q^2$-dependence, we use the model-independent $z$-expansion \cite{Hill2010},
\begin{align}
    \label{eq:zexp_GE}
    G_E^{p,n}(Q^2) &= \sum_{k=0}^m a_k^{p,n} z(Q^2)^k , \\
    \label{eq:zexp_GM}
    G_M^{p,n}(Q^2) &= \sum_{k=0}^m b_k^{p,n} z(Q^2)^k ,
\end{align}
with
\begin{equation}
    z(Q^2) = \frac{\sqrt{\tau_\mathrm{cut} + Q^2} - \sqrt{\tau_\mathrm{cut} - \tau_0}}{\sqrt{\tau_\mathrm{cut} + Q^2} + \sqrt{\tau_\mathrm{cut} - \tau_0}} ,
    \label{eq:z}
\end{equation}
where we employ $\tau_0 = 0$ and $\tau_\mathrm{cut} = 4M_{\pi, \mathrm{phys}}^2$.
We truncate the $z$-expansion beyond $m = 9$, and incorporate the four sum rules from Ref.\@ \cite{Lee2015} for each form factor, which ensure the correct asymptotic behavior of the latter for large $Q^2$.
The normalization of the electric form factor is enforced by fixing $a_0^p = 1$ and $a_0^n = 0$, respectively.
For the determination of the Zemach radius, we fit $G_E$ and $G_M$ simultaneously, similar to the crosscheck of our analysis in Ref.\@ \cite{Djukanovic2023}, so that we have eleven independent fit parameters altogether.
For the third Zemach moment, on the other hand, only the electric form factor is required, so that we fit only $G_E$ and have five independent fit parameters here.
The extrapolation fits are performed for the proton and neutron independently.
Using more than twenty data points for each form factor or a higher degree of the $z$-expansion does not increase the overlap between the original B$\chi$PT fit and the extrapolation any further.

For the numerical integration of \cref{eq:third_Zemach_moment,eq:Zemach_radius}, we smoothly replace the B$\chi$PT parametrization of the form factors by the $z$-expansion-based extrapolation in a narrow window around $Q^2_\mathrm{cut}$.
Concretely, we use the following estimate for the form factor term,
\begin{equation}
    F(Q^2) = \frac{1}{2}\left[ 1 - \tanh\left(\frac{Q^2 - Q^2_\mathrm{cut}}{\Delta Q^2_w}\right) \right] F^\chi(Q^2) + \frac{1}{2}\left[ 1 + \tanh\left(\frac{Q^2 - Q^2_\mathrm{cut}}{\Delta Q^2_w}\right) \right] F^z(Q^2) ,
    \label{eq:window}
\end{equation}
where $F(Q^2) \equiv G_E(Q^2) G_M(Q^2) / \mu_M$ for the Zemach radius and $F(Q^2) \equiv G_E^2(Q^2)$ for the third Zemach moment, respectively.
In \cref{eq:window}, $F^\chi(Q^2)$ represents our fit to B$\chi$PT, while $F^z(Q^2)$ denotes the $z$-expansion parametrization of the form factors.
For the width of the window in which we switch between the two parametrizations, we choose $\Delta Q^2_w = 0.0537 t_0^{-1} \approx 0.1~\mathrm{GeV}^2$.
We remark that for a consistent calculation of the third Zemach moment, the replacement according to \cref{eq:window} has to be applied to all terms in \cref{eq:third_Zemach_moment}, \ie also to the value of $\langle r_E^2 \rangle$.
The cancellation between the different terms of \cref{eq:third_Zemach_moment} at small $Q^2$ does not occur at the required numerical accuracy on all our bootstrap samples.
To facilitate the numerical integration, we therefore regulate the small-$Q^2$ contribution to the integral for the proton by replacing $t_0 Q^2 \to t_0 Q^2 + \num{1e-7}$ in the denominator, which changes the result for $\langle r_E^3 \rangle_{(2)}^p$ by less than 10~\% of its statistical error.

\begin{figure}[htb]
    \includegraphics[width=0.5\textwidth]{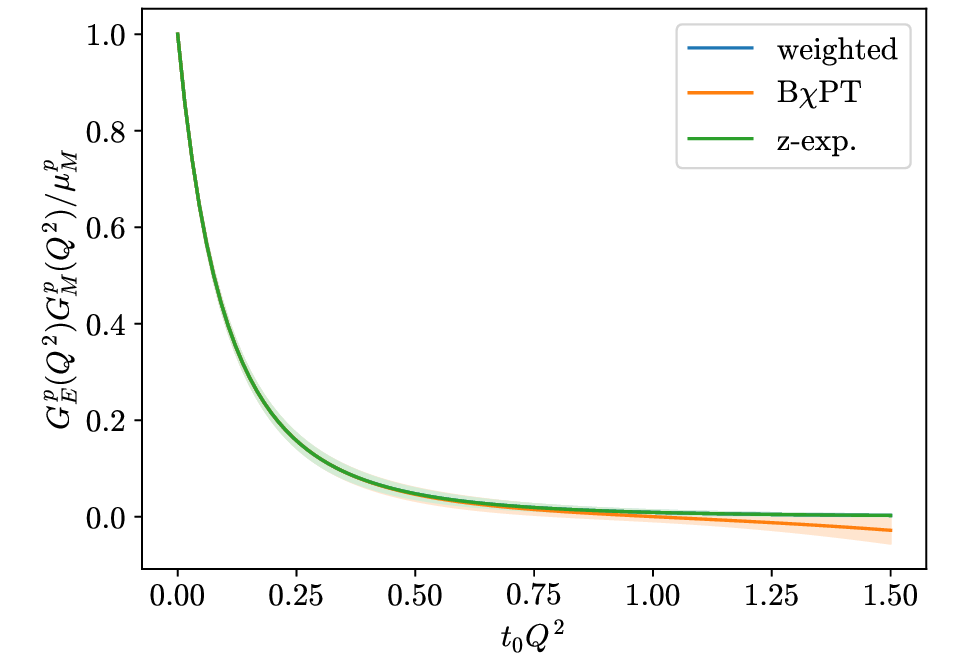}
    \caption{Product of the electric and normalized magnetic form factors of the proton at the physical point evaluated with different parametrizations. The orange curve shows one of the B$\chi$PT fits to our lattice data with $Q^2_\mathrm{cut} = 0.6~\mathrm{GeV}^2 \approx 0.322t_0^{-1}$, the green curve the $z$-expansion-based extrapolation, and the blue curve the weighted average of the two according to \cref{eq:window}.}
    \label{fig:formfactors}
\end{figure}

The two parametrizations and their weighted average according to \cref{eq:window} are illustrated in \cref{fig:formfactors} for the case of the Zemach radius of the proton.
While the B$\chi$PT formula is clearly not reliable for $Q^2 \gtrsim 1.7~\mathrm{GeV}^2 \approx 0.9t_0^{-1}$, the $z$-expansion behaves well for arbitrarily large momenta, which is due to the sum rules \cite{Lee2015} we have included.
In the region where we adjust the $z$-expansion to the B$\chi$PT parametrization ($0 < Q^2 \leq 0.6~\mathrm{GeV}^2$ for the case shown in \cref{fig:formfactors}), however, the two curves overlap so closely that they are indistinguishable by eye.
The blue curve, which is the one we use for the integration, smoothly switches from the orange (B$\chi$PT) curve to the green ($z$-expansion) one in a tight window around $Q^2_\mathrm{cut} = 0.6~\mathrm{GeV}^2 = 0.322t_0^{-1}$.

Replacing the B$\chi$PT parametrization smoothly with a constant zero instead of the $z$-expansion-based extrapolation [\ie setting $F^z(Q^2) \equiv 0$ in \cref{eq:window}] allows one to estimate the contribution of the form factors at $Q^2 > Q^2_\mathrm{cut}$ to the resulting Zemach radius and third Zemach moment, respectively.
For $Q^2_\mathrm{cut} = 0.6~\mathrm{GeV}^2$ (our largest, \ie least stringent, value for the cut), we find that the relative difference of the thus obtained value for $r_Z^p$ to the actual result using the corresponding variation of the B$\chi$PT fits is less than 0.9~\%.
In other words, the form factor term at $Q^2 > 0.6~\mathrm{GeV}^2$ contributes less than 0.9~\% to the Zemach radius of the proton.
For the third Zemach moment, the denominator in the integrand suppresses the large-$Q^2$ contribution to the integral even more strongly than for the Zemach radius.
Accordingly, we find a corresponding relative contribution of less than 0.3~\% to the third Zemach moment of the proton.

Due to this smallness of the contribution of the extrapolated form factors, the precise form of the chosen model for the extrapolation only has a marginal influence on the resulting values for the Zemach radius and third Zemach moment.
For example, if we replace the $z$-expansion by a dipole \ansatz (which also fulfills the constraints from Ref.\@ \cite{Lepage1980}), we find that the Zemach radius of the proton derived from any of our fit variations changes by at most 20~\% of the entire systematic error quoted in \cref{eq:result_rZp} below.
Thus, adding the variation in $r_Z^p$ due to the extrapolation model quadratically to the systematic uncertainty in \cref{eq:result_rZp} would not change the latter significantly.

Finally, we note that the major advantage of our approach based on the B$\chi$PT fits over an integration of the form factors on each ensemble is that the Zemach and Friar radii can be computed directly at the physical point, so that an extrapolation of results for the radii to the physical point, which would entail further significant systematic uncertainties, is not required.

%% file: content/model_average.tex
\section{Model average and final result}
\label{sec:model_average}
As in Refs.\@ \cite{Djukanovic2023,Djukanovic2023a}, we do not have a strong \apriori preference for one specific setup of the B$\chi$PT fits.
Thus, we determine our final results as well as the statistical and systematic error estimates from an average over the different fit models and kinematic cuts, using weights derived from the Akaike Information Criterion (AIC) \cite{Akaike1973,Akaike1974,Neil2024,Burnham2004,Borsanyi2015,Borsanyi2021}.
All values for the Zemach radii and third Zemach moments entering the average are listed in the \hyperref[sec:appendix]{Supplemental Material}, together with the associated weights.
More details on our procedure can be found in section V of Ref.\@ \cite{Djukanovic2023}.
As our final results, we obtain
\begin{align}
    \label{eq:result_rZp}
    r_Z^p &= ( 1.013 \pm 0.010\ (\mathrm{stat}) \pm 0.012\ (\mathrm{syst}) )~\mathrm{fm} , \\
    \label{eq:result_rF3p}
    \langle r_E^3 \rangle_{(2)}^p &= ( 2.200 \pm 0.060\ (\mathrm{stat}) \pm 0.071\ (\mathrm{syst}) )~\mathrm{fm}^3 , \\
    \label{eq:result_rZn}
    r_Z^n &= ( -0.0411 \pm 0.0056\ (\mathrm{stat}) \pm 0.0040\ (\mathrm{syst}) )~\mathrm{fm} , \\
    \label{eq:result_rF3n}
    \langle r_E^3 \rangle_{(2)}^n &= ( 0.0078 \pm 0.0020\ (\mathrm{stat}) \pm 0.0012\ (\mathrm{syst}) )~\mathrm{fm}^3 .
\end{align}
This corresponds to Friar radii of $r_F^p = ( 1.301 \pm 0.012\ (\mathrm{stat}) \pm 0.014\ (\mathrm{syst}) )~\mathrm{fm}$ and $r_F^n = ( 0.198 \pm 0.017\ (\mathrm{stat}) \pm 0.010\ (\mathrm{syst}) )~\mathrm{fm}$, respectively.

\begin{figure}[htb]
    \includegraphics[width=0.5\textwidth]{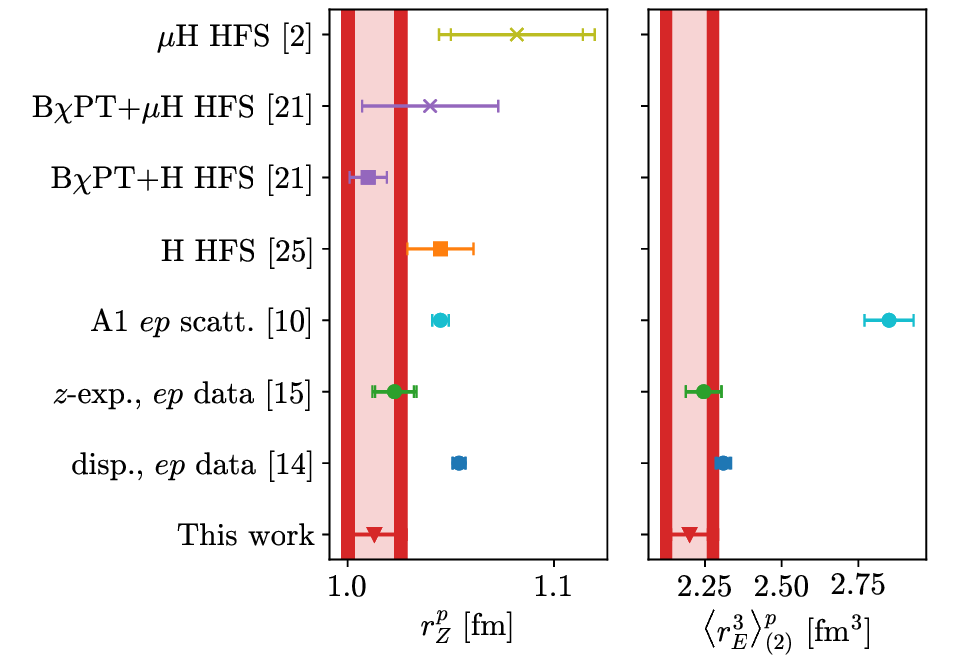}
    \caption{Comparison of our best estimates for the Zemach radius and third Zemach moment of the proton (red downward-pointing triangles) with determinations based on experimental data, \ie muonic hydrogen HFS \cite{Antognini2013,Hagelstein2023} (crosses), electronic hydrogen HFS \cite{Volotka2005,Hagelstein2023} (squares), and $ep$ scattering \cite{Distler2011,Borah2020,Lin2022} (circles).}
    \label{fig:comparison}
\end{figure}

In \cref{fig:comparison}, our numbers for the proton are compared to other determinations based on experimental data.
There are three main types of experiments which have been employed in the literature to compute the Zemach radius of the proton:
muonic hydrogen HFS \cite{Antognini2013}, electronic hydrogen HFS \cite{Hellwig1970}, and $ep$ scattering.
In order to extract the proton Zemach radius from an HFS measurement, input on the proton-polarizability effect is required.
This can be either taken from B$\chi$PT \cite{Hagelstein2023} or evaluated in a data-driven fashion, \ie using information on the spin structure functions \cite{Faustov2002,Cherednikova2002,Carlson2011} (as was done in Refs.\@ \cite{Volotka2005,Antognini2013}).
The form factors measured in $ep$-scattering experiments, on the other hand, can be analyzed with many different fit models, \eg by employing a (modified) power series \cite{Distler2011}, a $z$-expansion \cite{Borah2020}, or dispersion theory \cite{Lin2022}.

While our result for $r_Z^p$ agrees within one combined standard deviation with the extractions based on B$\chi$PT \cite{Hagelstein2023} and the $z$-expansion-based analysis of world $ep$-scattering data \cite{Borah2020}, and still within two combined standard deviations with the data-driven HFS extractions \cite{Volotka2005,Antognini2013} and the analysis of the A1 $ep$-scattering experiment \cite{Distler2011}, we observe a $2.6\,\sigma$ tension with the dispersive analysis of world $ep$-scattering data \cite{Lin2022}.
We also note that our estimate is smaller than all of the above experimental determinations except the one combining B$\chi$PT and electronic hydrogen HFS, which is slightly smaller than ours.

The proton's third Zemach moment can be extracted from $ep$-scattering experiments in the same way as its Zemach radius, and we also compare to these results in \cref{fig:comparison}.
Again, our value is comparatively small, but this time in good agreement with both the $z$-expansion-based \cite{Borah2020} and dispersive analyses \cite{Lin2022}.
Against the analysis of the A1 $ep$-scattering experiment \cite{Distler2011}, on the other hand, we observe a clear tension of $5.3\,\sigma$ in $\langle r_E^3 \rangle_{(2)}^p$.

In interpreting the aforementioned discrepancies, one must take into account that our results for the Zemach radii and third Zemach moments are not independent from those for the electromagnetic radii \cite{Djukanovic2023,Djukanovic2023a} because they are based on the same lattice data for the form factors and the same B$\chi$PT fits.
Indeed, we observe a correlation of around 80~\% both between $\sqrt{\langle r_E^2 \rangle^p}$ and $r_Z^p$ and between $\sqrt{\langle r_M^2 \rangle^p}$ and $r_Z^p$, while our correlation between $\sqrt{\langle r_E^2 \rangle^p}$ and $r_F^p$ is even larger, about 95~\%.
A large positive correlation between the proton's electric and Zemach radii has also been reported in the experimental literature \cite{Friar2005,Antognini2022}.
Hence, our small results for $\sqrt{\langle r_E^2 \rangle^p}$ and $\sqrt{\langle r_M^2 \rangle^p}$ in Refs.\@ \cite{Djukanovic2023,Djukanovic2023a} naturally imply small values for $r_Z^p$ and $r_F^p$.
By contrast, the dispersive analysis \cite{Lin2022} arrives at a significantly larger magnetic radius than the A1-data analyses \cite{Bernauer2014,Lee2015} and our lattice-QCD-based extraction \cite{Djukanovic2023,Djukanovic2023a}.
This may explain why we observe a larger tension in the Zemach radius (which equally probes electric and magnetic properties) with Ref.\@ \cite{Lin2022} than with Ref.\@ \cite{Distler2011}, even though the situation is exactly the opposite for the third Zemach moment / Friar radius (which only probes the electric properties).
For a deeper understanding of the underlying differences, a comparison of the full $Q^2$-dependence of the form factors would be required, rather than merely of the radii.
Furthermore, the role of higher-order electromagnetic corrections should be clarified.

The Zemach radius of the proton can also be computed in the framework of heavy-baryon chiral perturbation theory \cite{Peset2014}, which yields a much larger value of $r_Z^p = 1.35~\mathrm{fm}$.
However, the authors of Ref.\@ \cite{Peset2014} do not quote an error estimate on this number and claim it to be in good agreement with the experimental results, so that the uncertainty is presumably rather large.

Our results for the neutron are very well compatible with the $z$-expansion-based analysis of world $en$-scattering data \cite{Borah2020}, albeit with a more than 2 times larger error.

%% file: content/conclusions.tex
\section{Conclusions}
We have performed the first lattice-QCD calculation of the Zemach and Friar radii of the proton and neutron, which includes the contributions from quark-connected and -disconnected diagrams and presents a full error budget.
The overall precision of our results for the proton is sufficient to make a meaningful comparison to data-driven evaluations.
Our final estimates, which are given in \cref{eq:result_rZp,eq:result_rF3p,eq:result_rZn,eq:result_rF3n}, point to small values for the Zemach and Friar radii of the proton, but are consistent with most of the previous determinations within two standard deviations.
We agree rather well with the dispersive analysis of Ref.\@ \cite{Lin2022} regarding the electric properties of the proton (\ie the Friar radius), but to a much lesser degree on its magnetic properties (\ie the Zemach radius).

We stress that our results are highly correlated with those for the electromagnetic radii \cite{Djukanovic2023,Djukanovic2023a}.
Thus, our relatively low values for the Zemach and Friar radii of the proton are not unexpected, and they do not give rise to an independent puzzle from the lattice perspective.

%% file: content/acknowledgments.tex
\begin{acknowledgments}
    The authors thank Franziska Hagelstein for useful discussions.
    This research is partly supported by the Deutsche Forschungsgemeinschaft (DFG, German Research Foundation) through project HI 2048/1-2 (project No.\@ 399400745) and through the Cluster of Excellence \enquote{Precision Physics, Fundamental Interactions and Structure of Matter} (PRISMA${}^+$ EXC 2118/1) funded within the German Excellence Strategy (project ID 39083149).
    Calculations for this project were partly performed on the HPC clusters \enquote{Clover} and \enquote{HIMster2} at the Helmholtz Institute Mainz, and partly using the supercomputer \enquote{Mogon 2} operated by Johannes Gutenberg University Mainz (\url{https://hpc.uni-mainz.de}), which is a member of the AHRP (Alliance for High Performance Computing in Rhineland Palatinate, \url{https://www.ahrp.info}), the Gauss Alliance e.V., and the NHR Alliance (Nationales Hochleistungsrechnen, \url{https://www.nhr-verein.de}).
    The authors also gratefully acknowledge the John von Neumann Institute for Computing (NIC) and the Gauss Centre for Supercomputing e.V. (\url{https://www.gauss-centre.eu}) for funding this project by providing computing time on the GCS Supercomputer JUWELS at Jülich Supercomputing Centre (JSC) through projects CHMZ21, CHMZ36, NUCSTRUCLFL, and GCSNUCL2PT.

    Our programs use the QDP++ library \cite{Edwards2005} and the deflated SAP+GCR solver from the openQCD package \cite{Luescher2013}.
    The contractions have been explicitly checked using the Quark Contraction Tool \cite{Djukanovic2020}.
    We thank Simon Kuberski for providing the improved reweighting factors \cite{Kuberski2024} for the gauge ensembles used in our calculation.
    Moreover, we are grateful to our colleagues in the CLS initiative for sharing the gauge field configurations on which this work is based.
\end{acknowledgments}

%% file: content/appendix.tex
\section{Supplemental Material}
\label{sec:appendix}
Here, we present the results for the Zemach radii and third Zemach moments of the proton and neutron obtained from all our variations of the B$\chi$PT fits and the corresponding $z$-expansion-based extrapolations.
We apply different cuts in the pion mass ($M_\pi \leq 0.23~\mathrm{GeV}$ and $M_\pi \leq 0.27~\mathrm{GeV}$) and the momentum transfer ($Q^2 \leq 0.3, \ldots, 0.6~\mathrm{GeV}^2$).
The entries with and without an asterisk in the third column refer to a multiplicative and an additive model for the lattice-spacing ($a^2$) and/or finite-volume ($M_\pi L$) effects, respectively.
All variations which are presented here are included in our model average, with weights as given in the last column.
For further details on the B$\chi$PT fits and the correction models, we refer to section IV A of Ref.\@ \cite{Djukanovic2023}, while the $\overline{\mathrm{BAIC}}$ weights are defined in section V therein.

\begin{longtable*}[c]{cccllccl}
    \caption{Results for the Zemach radii and third Zemach moments of the proton and neutron} \\*
    \hline\hline \\* [.01em]
    $M_{\pi, \mathrm{cut}}$ [GeV] & $Q^2_\mathrm{cut}$ [GeV${}^2$] & correction             & \multicolumn{1}{c}{$r_Z^p$ [fm]} & \multicolumn{1}{c}{$\langle r_E^3 \rangle_{(2)}^p$ [fm${}^3$]} & $r_Z^n$ [fm] & $\langle r_E^3 \rangle_{(2)}^n$ [fm${}^3$] & \multicolumn{1}{c}{$\overline{\mathrm{BAIC}}$ weight} \\* [.05em] \hline \\* [.05em]
    \endfirsthead
    \caption[]{\emph{(Continued)}} \\*
    \hline\hline \\* [.01em]
    $M_{\pi, \mathrm{cut}}$ [GeV] & $Q^2_\mathrm{cut}$ [GeV${}^2$] & correction             & \multicolumn{1}{c}{$r_Z^p$ [fm]} & \multicolumn{1}{c}{$\langle r_E^3 \rangle_{(2)}^p$ [fm${}^3$]} & $r_Z^n$ [fm] & $\langle r_E^3 \rangle_{(2)}^n$ [fm${}^3$] & \multicolumn{1}{c}{$\overline{\mathrm{BAIC}}$ weight} \\* [.05em] \hline \\* [.05em]
    \endhead
    \hline
    \endfoot
    \hline\hline
    \endlastfoot
    0.23                          & 0.3                            & --                     & 1.0153(97) & 2.202(72) & -0.0423(39) & 0.0080(15) & \num{0.0597} \\
    0.23                          & 0.3                            & $a^2$                  & 0.999(15)  & 2.135(83) & -0.0345(64) & 0.0058(18) & \num{0.0138} \\
    0.23                          & 0.3                            & ${}^*a^2$              & 0.994(15)  & 2.06(10)  & -0.0351(70) & 0.0058(21) & \num{0.0390} \\
    0.23                          & 0.3                            & $M_\pi L$              & 1.016(11)  & 2.195(73) & -0.0427(41) & 0.0081(16) & \num{0.00135} \\
    0.23                          & 0.3                            & ${}^*M_\pi L$          & 1.017(12)  & 2.200(78) & -0.0435(43) & 0.0084(17) & \num{0.00117} \\
    0.23                          & 0.3                            & $a^2, M_\pi L$         & 0.998(18)  & 2.118(87) & -0.0337(68) & 0.0056(19) & \num{0.000279} \\
    0.23                          & 0.3                            & ${}^*a^2, {}^*M_\pi L$ & 0.993(18)  & 2.04(11)  & -0.0355(77) & 0.0059(24) & \num{0.000586} \\
    [\defaultaddspace]
    0.23                          & 0.4                            & --                     & 1.0209(81) & 2.261(52) & -0.0421(48) & 0.0080(18) & \num{0.0198} \\
    0.23                          & 0.4                            & $a^2$                  & 1.003(13)  & 2.183(60) & -0.0348(67) & 0.0060(19) & \num{0.0611} \\
    0.23                          & 0.4                            & ${}^*a^2$              & 0.996(13)  & 2.097(79) & -0.0342(76) & 0.0056(23) & \num{0.129} \\
    0.23                          & 0.4                            & $M_\pi L$              & 1.0213(99) & 2.253(52) & -0.0430(53) & 0.0083(20) & \num{0.00106} \\
    0.23                          & 0.4                            & ${}^*M_\pi L$          & 1.022(11)  & 2.263(58) & -0.0440(56) & 0.0087(22) & \num{0.000845} \\
    0.23                          & 0.4                            & $a^2, M_\pi L$         & 1.001(17)  & 2.162(64) & -0.0337(75) & 0.0057(21) & \num{0.00141} \\
    0.23                          & 0.4                            & ${}^*a^2, {}^*M_\pi L$ & 0.994(16)  & 2.076(92) & -0.0346(86) & 0.0058(26) & \num{0.00180} \\
    [\defaultaddspace]
    0.23                          & 0.5                            & --                     & 1.0275(74) & 2.279(43) & -0.0427(48) & 0.0084(19) & \num{0.0840} \\
    0.23                          & 0.5                            & $a^2$                  & 1.015(12)  & 2.239(44) & -0.0401(66) & 0.0077(22) & \num{0.0354} \\
    0.23                          & 0.5                            & ${}^*a^2$              & 1.010(11)  & 2.165(62) & -0.0397(76) & 0.0075(27) & \num{0.0867} \\
    0.23                          & 0.5                            & $M_\pi L$              & 1.0283(96) & 2.265(45) & -0.0432(50) & 0.0086(19) & \num{0.00402} \\
    0.23                          & 0.5                            & ${}^*M_\pi L$          & 1.030(10)  & 2.276(52) & -0.0442(53) & 0.0089(21) & \num{0.00314} \\
    0.23                          & 0.5                            & $a^2, M_\pi L$         & 1.014(16)  & 2.213(51) & -0.0389(70) & 0.0074(23) & \num{0.00101} \\
    0.23                          & 0.5                            & ${}^*a^2, {}^*M_\pi L$ & 1.009(15)  & 2.140(80) & -0.0397(82) & 0.0076(29) & \num{0.00159} \\
    [\defaultaddspace]
    0.23                          & 0.6                            & --                     & 1.0272(75) & 2.280(45) & -0.0424(45) & 0.0084(17) & \num{0.103} \\
    0.23                          & 0.6                            & $a^2$                  & 1.014(12)  & 2.236(47) & -0.0389(64) & 0.0074(21) & \num{0.0536} \\
    0.23                          & 0.6                            & ${}^*a^2$              & 1.009(11)  & 2.165(62) & -0.0378(76) & 0.0070(25) & \num{0.0731} \\
    0.23                          & 0.6                            & $M_\pi L$              & 1.0261(96) & 2.271(47) & -0.0432(49) & 0.0086(18) & \num{0.00469} \\
    0.23                          & 0.6                            & ${}^*M_\pi L$          & 1.028(10)  & 2.282(54) & -0.0443(51) & 0.0090(20) & \num{0.00412} \\
    0.23                          & 0.6                            & $a^2, M_\pi L$         & 1.012(16)  & 2.216(52) & -0.0378(71) & 0.0071(22) & \num{0.000887} \\
    0.23                          & 0.6                            & ${}^*a^2, {}^*M_\pi L$ & 1.008(15)  & 2.148(80) & -0.0381(82) & 0.0071(27) & \num{0.000941} \\
    [\defaultaddspace]
    0.27                          & 0.3                            & --                     & 1.0149(89) & 2.186(70) & -0.0435(36) & 0.0083(14) & \num{0.110} \\
    0.27                          & 0.3                            & $a^2$                  & 1.002(13)  & 2.136(77) & -0.0436(58) & 0.0083(21) & \num{0.00900} \\
    0.27                          & 0.3                            & ${}^*a^2$              & 0.999(13)  & 2.080(93) & -0.0447(64) & 0.0088(25) & \num{0.0123} \\
    0.27                          & 0.3                            & $M_\pi L$              & 1.016(11)  & 2.179(72) & -0.0433(38) & 0.0082(15) & \num{0.00122} \\
    0.27                          & 0.3                            & ${}^*M_\pi L$          & 1.017(11)  & 2.186(76) & -0.0437(40) & 0.0084(16) & \num{0.00121} \\
    0.27                          & 0.3                            & $a^2, M_\pi L$         & 1.001(16)  & 2.120(81) & -0.0429(62) & 0.0081(22) & \num{9.86e-05} \\
    0.27                          & 0.3                            & ${}^*a^2, {}^*M_\pi L$ & 0.998(16)  & 2.06(11)  & -0.0450(71) & 0.0089(28) & \num{0.000110} \\
    [\defaultaddspace]
    0.27                          & 0.4                            & --                     & 1.0205(74) & 2.249(50) & -0.0434(44) & 0.0085(17) & \num{0.0225} \\
    0.27                          & 0.4                            & $a^2$                  & 1.006(12)  & 2.198(56) & -0.0421(60) & 0.0081(21) & \num{0.00509} \\
    0.27                          & 0.4                            & ${}^*a^2$              & 1.002(11)  & 2.136(72) & -0.0432(68) & 0.0084(25) & \num{0.00682} \\
    0.27                          & 0.4                            & $M_\pi L$              & 1.0211(94) & 2.242(51) & -0.0435(50) & 0.0085(19) & \num{0.000422} \\
    0.27                          & 0.4                            & ${}^*M_\pi L$          & 1.022(10)  & 2.252(56) & -0.0443(53) & 0.0088(21) & \num{0.000350} \\
    0.27                          & 0.4                            & $a^2, M_\pi L$         & 1.004(15)  & 2.179(60) & -0.0416(68) & 0.0079(23) & \num{6.60e-05} \\
    0.27                          & 0.4                            & ${}^*a^2, {}^*M_\pi L$ & 1.000(15)  & 2.119(86) & -0.0441(78) & 0.0088(30) & \num{5.45e-05} \\
    [\defaultaddspace]
    0.27                          & 0.5                            & --                     & 1.0259(67) & 2.276(39) & -0.0439(44) & 0.0088(18) & \num{0.0253} \\
    0.27                          & 0.5                            & $a^2$                  & 1.011(11)  & 2.244(41) & -0.0437(58) & 0.0087(21) & \num{0.00361} \\
    0.27                          & 0.5                            & ${}^*a^2$              & 1.010(10)  & 2.188(58) & -0.0454(67) & 0.0094(26) & \num{0.00416} \\
    0.27                          & 0.5                            & $M_\pi L$              & 1.0282(91) & 2.262(42) & -0.0438(48) & 0.0088(19) & \num{0.000499} \\
    0.27                          & 0.5                            & ${}^*M_\pi L$          & 1.0294(98) & 2.275(49) & -0.0444(51) & 0.0091(20) & \num{0.000410} \\
    0.27                          & 0.5                            & $a^2, M_\pi L$         & 1.011(14)  & 2.218(47) & -0.0430(64) & 0.0085(23) & \num{6.11e-05} \\
    0.27                          & 0.5                            & ${}^*a^2, {}^*M_\pi L$ & 1.010(14)  & 2.167(75) & -0.0459(73) & 0.0097(29) & \num{4.02e-05} \\
    [\defaultaddspace]
    0.27                          & 0.6                            & --                     & 1.0246(69) & 2.284(43) & -0.0435(42) & 0.0087(16) & \num{0.00555} \\
    0.27                          & 0.6                            & $a^2$                  & 1.007(11)  & 2.243(43) & -0.0426(64) & 0.0084(21) & \num{0.00246} \\
    0.27                          & 0.6                            & ${}^*a^2$              & 1.008(11)  & 2.190(57) & -0.0440(74) & 0.0089(27) & \num{0.00119} \\
    0.27                          & 0.6                            & $M_\pi L$              & 1.0265(93) & 2.277(45) & -0.0435(46) & 0.0087(17) & \num{8.66e-05} \\
    0.27                          & 0.6                            & ${}^*M_\pi L$          & 1.028(10)  & 2.289(53) & -0.0444(49) & 0.0091(19) & \num{7.33e-05} \\
    0.27                          & 0.6                            & $a^2, M_\pi L$         & 1.006(15)  & 2.223(49) & -0.0419(72) & 0.0082(23) & \num{2.29e-05} \\
    0.27                          & 0.6                            & ${}^*a^2, {}^*M_\pi L$ & 1.007(14)  & 2.176(75) & -0.0447(82) & 0.0092(31) & \num{7.54e-06}
\end{longtable*}